\documentclass[onecolumn,showpacs]{revtex4}

\usepackage{graphicx}
\usepackage{dcolumn}
\usepackage{bm}

\input epsf

\begin{document}

\title{\Large  FRW Universe in Ho\~{r}ava Gravity}

\author{\bf~Nairwita~Mazumder\footnote{nairwita15@gmail.com}, Subenoy~Chakraborty\footnote{schakraborty@math.jdvu.ac.in}.}

\affiliation{$^1$Department of Mathematics,~Jadavpur
University,~Kolkata-32, India.}

\date{\today}

\begin{abstract}
Recently, a field theoretic model for a UV complete theory of
gravity has been proposed by Ho\~{r}ava. This theory is a
non-relativistic renormalizable gravity theory which coincides
with Einstein's general relativity at large distances.
Subsequently L\"{u} et al have formulated the modified Friedmann
equations and have presented a solution in vacuum. In the present
work, we rewrite the modified FRW equations in the form of usual
FRW equations in Einstein gravity and consequences has been
analyzed. Also the thermodynamics of the FRW universe has been
studied.\\


\end{abstract}


\maketitle

\section{\normalsize\bf{Introduction}}

Recently, a renormalizable theory of gravity was proposed by
Ho\~{r}ava [1,2]. As this theory follows Lifshitz-type
anisotropic scaling so it is commonly known as
Ho\~{r}ava-Lifshitz (HL) gravity. This theory of gravitation has
four possible versions so far - with/without the detailed balance
condition and with/without projectability condition. Among these
version without detailed balance and with the the projectability
condition is the most viable one.\\
 Due to detailed balance condition the potential in the $4D$
Lorentzian action has a specific form in terms of a $3D$
Euclidean theory and it leads to obstacles from cosmological view
point. The projectability condition on the otherhand, is due to
the foliation preserving diffeomorphism invariance - the
fundamental symmetry of the theory. This new theory of gravity do
not considered the space and time on an equal footing- only the
general covariance is retained at large distance and coincides
with general relativity. In fact, it is a non-relativistic
renormalizable field theory model for gravity having UV
completeness. Using parameterized form [2] of the four
dimensional metric as

\begin{equation}
ds^{2}=-N^{2}~dt^{2}~+~g_{ij}(dx^{i}-N^{i}dt)(dx^{j}-N^{j}dt)
\end{equation}

and using ADM formalism, the Einstein-Hilbert action has the form

\begin{equation}
S_{EH}=\frac{1}{16\pi G}\int~d^{4}x
\sqrt{g}~N(k_{ij}k^{ij}-k^{2}+R-2\Lambda)~,
\end{equation}

where $G$ is the Newton's constant and $k_{ij}$, the extrinsic
curvature of a space-like hyper surface with a fixed time has the
expression

\begin{equation}
k_{ij}=~\frac{1}{2N}~[g_{ij}- (\nabla_{i} N_{j})(\nabla_{j}N_{i})]
\end{equation}

Here a over dot denotes derivative with respect to $'t'$ and
covariant derivatives are defined with respect to the spatial
metric $g_{ij}$.\\

The action of the proposed non-relativistic generalized theory by
Ho\~{r}ava [1] has the expression (knowwn as Ho\~{r}ava -Lifshitz
(HL) action)

\begin{equation}
S_{HL}=\int~dt d^{3}x \sqrt{g}~N(L_{0}+L_{1})
\end{equation}

with

$$L_{0}= \left[\frac{2}{k^{2}}(k_{ij}k^{ij}-\lambda k^{2})+ ~\frac{k^{2} \mu^{2}(\Lambda
R-3\Lambda^{2})}{8(1-3\lambda)}\right]$$ and

\begin{equation}
L_{1}=\left[\frac{k^{2} \mu^{2}
(1-4\lambda)R^{2}}{32(1-3\lambda)}-\frac{k^{2}z_{ij}z^{ij}}{2\omega^4}\right]
\end{equation}

Here $$z_{ij}=c_{ij}-\frac{\mu \omega^{2}}2 R_{ij}~,$$

\begin{equation}
c^{ij}=\epsilon^{ikl}
\nabla_{k}({R^{j}}_{l}-\frac{1}4{\delta^{j}}_{l}) =\epsilon^{ikl}
\nabla_{k}{R^{j}}_{l}-\frac{1}4\epsilon^{ikj} \partial_{k}R~;
\end{equation}

is known as Cotten tensor and $k^{2},\lambda,\mu,\omega,
\Lambda~$ are constant parameters.\\

In the above HL-action (4) the first two terms are kinetic terms
and the rest correspond to potential of the theory (in 'detailed
balance' form). A comparative study with general relativity shows
that the speed of light, Newton's constant and the cosmological
constant have the expressions

\begin{equation}
c=\frac{k^{2}\mu}4
\sqrt{\frac{\Lambda}{1-3\lambda}}~,~G=\frac{k^{2}c}{32 \pi}~,~
{\tilde \Lambda}=\frac{3}2\Lambda~.
\end{equation}

In the present theory [1] $\lambda$ is a dynamical coupling
constant, subject to quantum correction. For $\lambda=1$ the
first three terms in the action (4) are the usual one's in
Einstein's general relativity. Also from the equation (7) the
expression for the velocity of light shows that $\Lambda$ should
be negative if $\lambda>\frac{1}3$. Note that the HL-action (4)
remains real [3] under the analytic continuation [4]

\begin{equation}
\mu \rightarrow i \mu~,~\omega^{2} \rightarrow -i\omega^{2}~,
\end{equation}

so that one may choose $\Lambda$ to be positive for $\lambda >
\frac{1}3$. The cosmological implications of the HL-actions has
been studied in [5-11].\\

Now variation of the HL-action with respect to
$N,N^{i}$~and~$g_{ij}$ gives the equations of motion

\begin{equation}
\frac{2}{k^{2}}(k_{ij}k^{ij}-\lambda k^{2})-\frac{k^{2} \mu^{2}
(1-4\lambda) R^{2}}{32(1-3\lambda)}
+\frac{k^{2}z_{ij}z^{ij}}{2\omega^4}-~\frac{k^{2} \mu^{2}(\Lambda
R-3\Lambda^{2})}{8(1-3\lambda)}=0
\end{equation}

\begin{equation}
\nabla_{l} (k^{ls}-\lambda k g^{ls})=0
\end{equation}
and

\begin{equation}
\frac{2}{k^{2}}E_{ij}^{(1)}-\frac{2\lambda}{k^{2}} E_{ij}^{(2)} +
\frac{k^{2} \mu^{2} \Lambda}{8(1-3 \lambda)}E_{ij}^{(3)} +
\frac{k^{2} \mu^{2} (1-4\lambda)}{32(1-3\lambda)} E_{ij}^{(4)} -
\frac{\mu k^{2}}{4 \omega^{2}}E_{ij}^{(5)} - \frac{k^{2}}{2
\omega^{4}}E_{ij}^{(6)} = 0
\end{equation}

where the tensors ${E_{ij}}^{(\alpha)}~(\alpha=1,2,...,6)$ are
combination of $k_{ij},g_{ij},N,N_{i}$ and their covariant
derivatives with respect to the three dimensional metric and
detailed expressions can be found in [3].\\

After Ho\~{r}ava [1] developed the new gravity theory, within
two-three months L\"{u} et al [3] have obtained static,
spherically symmetric solutions in HL gravity and have shown
asymptotically  $AdS_{4}$ solution for $\lambda=1$. Also they have
formulated the modified FRW equations and have obtained vacuum
solution for the isotropic model.\\

In the present work, we rearrange the modified Friedmann equation
so that it can be written in the usual Friedmann equations and
interprets the extra terms from the point of view of cosmology.
Also thermodynamics of the FRW universe in HL theory will be
investigated.\\

\section{\normalsize\bf{Friedmann Equations in HL-Gravity}}

Immediately, after the proposal for the new gravity gravity
theory by Ho\~{r}va [1], L\"{u} et al [4] give cosmological
solutions for this theory. At first they have solved the equations
of motion for spherically symmetric space-time model and have
shown the correspondence with $AdS_{4}$ asymptotically for
$\lambda=1$. They have also obtained solution, deviating slightly
from the detailed balance by changing the lagrangian in (4) as
$$\textit{L}=L_{0}+ (1- \epsilon^{2})L_{1}.$$ They have found that
for $\epsilon\neq 0$ the metric has a finite mass which diverges
in the detailed-balance limit $(\epsilon=0)$. Further, they have
written down the Friedmann equations in the new gravity theory
and has solved these Friedmann equations for vacuum case. For
$k=+1$, they have obtained a
bounce in the solution.\\

For the Fridmann-Lema$\acute{i}$tre-Robertson-Walker model of the
space-time with line element
$$ds^{2}=-c^{2}dt^{2}+{a(t)}^{2}\left[\frac{dr^{2}}{1-kr^{2}}~+~r^{2}(d\theta^{2}~+~\sin^{2}\theta d
\phi^{2})\right]$$

($k=1,0,-1$ correspond to a closed, flat or open universe) the
non-vanishing equations of motion are [4]

\begin{equation}
{\left(\frac{\dot{a}}a\right)}^{2}= \frac{2c^{4}}{(3 \lambda -1)}
\left[\frac{\Lambda}2 + \frac{8 \pi G \rho}3 - \frac{k}{a^{2}} +
\frac{k^{2}}{2\Lambda ~a^{4}} \right]
\end{equation}

and

\begin{equation}
\left(\frac{\ddot{a}}a\right)= \frac{2c^{4}}{(3 \lambda -1)}
\left[\frac{\Lambda}2 - \frac{4 \pi G (\rho+3p)}3 -
\frac{k^{2}}{2\Lambda ~a^{4}} \right]
\end{equation}

or equivalently we write

\begin{equation}
H^{2}= \frac{2c^{4}}{(3 \lambda -1)} \left[\frac{\Lambda}2 +
\frac{\kappa_{4}^{2} \rho}3 - \frac{k}{a^{2}} +
\frac{k^{2}}{2\Lambda ~a^{4}} \right]
\end{equation}

and

\begin{equation}
\dot{H}= \frac{2c^{4}}{(3 \lambda -1)}
\left[\frac{-\kappa_{4}^{2} (\rho+p)}2 + \frac{k}{a^{2}} -
\frac{k^{2}}{\Lambda ~a^{4}} \right]
\end{equation}

where $H=\frac{\dot{a}}a$~, usual Hubble parameter ,
$\kappa_{4}^{2}=8 \pi G$ and $\rho , p$ are respectively the
thermodynamic energy density and pressure of the fluid in the
universe. Note that for $k=0$ , there is no contribution from the
higher order derivative terms in the action. However, for
$k\neq0$, these higher derivative terms are significant for small
volume (i.e. small a) and become insignificant for large a, where
it agrees with general relativity.\\

Now, choosing $\frac{2c^{4}}{3\lambda-1}=1$ the above Friedmann
equations can be written as

\begin{equation}
H^{2}+\frac{k}{a^{2}}=\frac{\kappa_{4}^{2}}3(\rho+\rho_{HL})
\end{equation}

and

\begin{equation}
\dot{H}-\frac{k}{a^{2}}=- \frac{\kappa_{4}^{2}}2
(\rho+p+\rho_{HL}+p_{HL})
\end{equation}

where

\begin{equation}
\rho_{HL}=\frac{3}{\kappa_{4}^{2}}\left(\frac{\Lambda}2 +
\frac{k^{2}}{2 \Lambda a^{4}}\right)
\end{equation}

and

\begin{equation}
p_{HL}=\frac{1}{\kappa_{4}^{2}}\left(-\frac{3 \Lambda}2 +
\frac{k^{2}}{2 \Lambda a^{4}}\right)
\end{equation}

One may note that, the equations (16) and (17) are same as the
usual Friedmann equations in Einstein gravity having two fluid
system - one the usual fluid present in the universe and the
other may be interpreted as the effect of the HL gravity. If we
write ~$\rho_{t}=\rho+\rho_{HL}$~and~$ p_{t}=p+p_{HL}$~, then from
equations (16) and (17) the conservation equation will be

\begin{equation}
\dot{\rho_{t}}+3H(\rho_{t}+p_{t})=0
\end{equation}

Now if we assume the energy conservation for the ordinary matter
i.e.
\begin{equation}
\dot{\rho}+3H(\rho+p)=0
\end{equation}

then combining (20) and (21) and using the definition of
$\rho_{t}$ and $p_{t}$ we have

\begin{equation}
\dot{\rho_{HL}}+3H(\rho_{HL}+p_{HL})=0
\end{equation}

Hence we may say that the apparent two fluid system are
non-interacting. Therefore we may conclude that in the present
cosmological setting gravity in HL theory may be considered as the
Einstein gravity with two
non-interacting fluid system.\\

We now study the induced fluid systems due to HL gravity. If the
cosmological constant is positive then $\rho_{HL} > 0$ throughout
the evolution while $p_{HL}$ is initially positive and becomes
negative at ~$a^{2}=\frac{|k|}{(\sqrt{3}) \Lambda }$~. On the
other hand, for negative cosmological constant $\rho_{HL}$ is
always negative while $p_{HL}$ starts with negative value but
becomes positive when ~$a^{2}=\frac{|k|}{(\sqrt{3}) |\Lambda| }$~.
However, for $k=0$, it behaves as a cosmological constant.
Finally, for large $'a'$ whatever be the choice of $k$ the effect
of HL gravity reduces to a cosmological constant i.e. HL-gravity
becomes Einstein gravity with a cosmological constant at large
$'a'$.\\

\section{\normalsize\bf{Thermodynamics of FRW Universe in HL-Gravity}}

It is well known in the literature that the laws of
thermodynamics are valid for the universe bounded by the apparent
horizon. This is true not only in Einstein gravity [12-14] but
also in higher derivative Lovelock theory [15] of gravity. As the
present HL gravity theory is shown to be the generalization of
Einstein gravity by including an effective matter term to the
original matter so it is expected that laws of thermodynamics
will be valid on the apparent horizon. In this section we shall
examine the validity of the generalized second law of
thermodynamics assuming the first law of thermodynamics on the
event horizon. Also the matter in the universe is chosen as the
holographic dark energy. Form the principle of the holographic
dark energy [16] model the matter density of the holographic dark
energy component can be written as [16]

\begin{equation}
\rho_{D}=3c^{2}{R_{E}}^{-2}
\end{equation}

where $c$~is any arbitrary parameter. Now the form of the
equations of motion are the following:

\begin{equation}
H^{2}+\frac{k}{a^{2}}=\frac{\kappa_{4}^{2}}3(\rho_{D}+\rho_{HL})
\end{equation}

and

\begin{equation}
\dot{H}-\frac{k}{a^{2}}=- \frac{\kappa_{4}^{2}}2
(\rho_{D}+p_{D}+\rho_{HL}+p_{HL})
\end{equation}

Using the definition of event horizon

\begin{equation}
{R}_{E}=a\int^{\infty}_{a}\frac{da}{Ha^{2}}=\frac{c}{(\sqrt{\Omega_{D}})H}
\end{equation}

where ~$\Omega_{D}=\frac{\rho_{D}}{3H^{2}}$~is the density
parameter corresponding to dark energy. The equation of state for
the dark energy can be written as

\begin{equation}
\rho_{D}=\omega_{D}p_{D}
\end{equation}

with~$\omega_{D}$~is not necessarily a constant.\\

The amount of energy crossing the event horizon in time $dt$~ is
given by the expression [15]

\begin{equation}
-dE=4\pi{R_{E}}^{3}H(\rho_{t}+p_{t})dt~
\end{equation}

Thus assuming the validity of the first law of thermodynamics the
time variation of the horizon entropy is given by

\begin{equation}
\frac{dS_{E}}{dt}=\frac{4\pi {R_{E}}^3H}{T_{E}}(\rho_{t}+p_{t})
\end{equation}

where $S_{E}$~and~$T_{E}$~are respectively the entropy and
temperature of the event horizon.\\

To determine the time variation of the entropy of the matter
inside the event horizon we use the Gibb's equation [17]

\begin{equation}
T_{E}dS_{I}=dE_{I}+p_{t}dV
\end{equation}

where ~$S_{I}$~ and ~$E_{I}$~ are the entropy and energy of the
matter inside the event horizon. Note that due to thermodynamical
equilibrium we choose the temperature of the matter distribution
is same as that of the boundary surface (the event horizon).\\

Using
$$E_{I}= \frac{4}3\pi {R_{E}}^{3}\rho_{t}~~and~~V=\frac{4}3\pi
{R_{E}}^{3}~,$$

in the Gibb's equation and with the help of equations of motion
(24) and (25) we have

\begin{equation}
dS_{I}=\frac{4\pi {R_{E}}^{2}}{T_{E}}(\rho_{t}+p_{t})dR_{E}+
\frac{H{R_{E}}^{3}}{T_{E}}(\dot{H}-\frac{k}{a^{2}})dt
\end{equation}

To obtain~$(dR_{E})$~we start with the expression of $R_{E}$ in
equation (26) and using the conservation equation (20) for
holographic dark energy, we obtain

\begin{equation}
dR_{E}=\frac{3}2R_{E}H(1+\omega_{D})dt
\end{equation}

Hence the time variation of the matter entropy is given by (after
some simplification )

\begin{equation}
\frac{dS_{I}}{dt}=\frac{2\pi
{R_{E}}^3}{T_{E}}H(\rho_{t}+p_{t})(3\omega_{D}+1)
\end{equation}

Thus combining equations (29) and (33) the resulting change of
total entropy is given by

\begin{equation}
\frac{d}{dt}(S_{I}+S_{E})=\frac{6\pi {R_{E}}^3H}{T_{E}}
(\rho_{t}+p_{t})(\omega_{D}+1)
\end{equation}

which gives the same form as in Einstein gravity [18]. But here
the restrictions are different from that of the Einstein
gravity.\\

Now using the deceleration parameter $q=-1-\frac{\dot{H}}{H^{2}}$
the above expression can be written as

\begin{equation}
\frac{d}{dt}(S_{I}+S_{E})=\frac{12\pi {R_{E}}^3H}{T_{E}}
\left[(1+q)H^{2}+\frac{k}{a^{2}}\right](\omega_{D}+1)
\end{equation}

Before going to examine the validity of the second law
(generalized) of thermodynamics we first write the explicit form
of~$(\rho_{t}+p_{t})$~as follows:

\begin{equation}
\rho_{t}+p_{t}=\rho_{D}+\frac{3}{\kappa_{4}^{2}}\left(\frac{\Lambda}2
+ \frac{k^{2}}{2 \Lambda a^{4}}\right) +
p_{D}+\frac{1}{\kappa_{4}^{2}}\left(-\frac{3 \Lambda}2 +
\frac{k^{2}}{2 \Lambda a^{4}}\right)~=~\rho_{D}(1+\omega_{D})+
\frac{2k^{2}}{\kappa_{4}^{2} \Lambda a^{4}}
\end{equation}

The conclusions are the following:\\

{\bf{case I: $\Lambda > 0$}}\\

If the holographic dark energy satisfies the weak energy
condition then the generalized second law of thermodynamics will
always be satisfied as in Einstein gravity. However, if the dark
energy does not obey the weak energy condition then the result is
distinct from Einstein gravity: at very early stages of the
evolution of the universe, HL term (i.e the $\ Lambda$ term) in
equation (36)dominates and there is violation of the second law
of thermodynamics. But at later epoch the HL term becomes
insignificant, the second law of thermodynamics will again be
valid. \\

{\bf{case II: $\Lambda < 0$}}\\

We see from (36) that $'\Lambda'$ term (i.e. HL term) may have a
significant contribution at very early stages of the evolution of
the universe. This means that if the holographic dark energy
satisfies the weak energy condition, $\rho_{t}+p_{t}$ may not be
positive at very early stages of the evolution of the universe.
Thus second law of thermodynamics may not be satisfied at early
stages of the evolution of the universe even if the dark energy
satisfies the weak energy condition. However at later stages of
the evolution when HL-term becomes insignificant, $\rho_{t}+p_{t}$
becomes positive then
generalized second law is obeyed.\\
Further, if the weak energy condition is not satisfied by the
dark energy (phantom in nature) then second law of thermodynamics
will always be satisfied.Hence we may conclude that HL term has a
significant effect for the validity of the second law of
thermodynamics compare
to Einstein gravity particularly when dark energy violates weak energy condition.\\

For future work, it will be interesting to examine the validity
of the first law of thermodynamics at the event horizon.\\

{\bf Acknowledgement:}\\

This paper has been carried out during a visit to IUCAA, Pune,
India.The authors are thankful to IUCAA for warm hospitality and
facility of doing research works.\\

{\bf References:}\\

$[1]$ P. Ho\~{r}ava  , {\it Phys. Rev. D} {\bf 79} 084008 (2009);\\

$[2]$ R. L. Arnowitt , S. Deser and C.W. Misner , The Dynamics of
General Relativity, " gravitation: an introduction to current
research , L. Witten ed."  (Wiley 1962), Chapter 7, pp 227-265,
arXiv: gr-qc/ 0405109;\\

$[3]$ G. Calcagni , arXiv: 0904.0829 [hep-th];\\

$[4]$ H. L\"{u} , J. Mei ,and C.N. Pope , arXiv: 0904.1595 [hep-th];\\

$[5]$ E. Kiritsis and G. Kofinas , arXiv: 0904.1334 [hep-th];\\

$[6]$ P. Ho\~{r}ava  , {\it JHEP } {\bf 0903} 020 (2009);\\

$[7]$ P. Ho\~{r}ava  , arXiv: 0902.3657 [hep-th];\\

$[8]$ T. Takahasi and J. Soda , arXiv: 0904.0554 [hep-th];\\

$[9]$ J. Klusan , arXiv: 0904.1343 [hep-th];\\

$[10]$ R.G. Cai , L.M. Cao and N. Ohta , arXiv: 0904.3670
[hep-th];\\

$[11]$ R.G. Cai , L.M. Cao and N. Ohta , arXiv: 0905.0751
[hep-th];\\

$[12]$ B. Wang, Y. Gong, E. Abdalla, \it{Phys. Rev. D} {\bf 74} 083520 (2006).\\

$[13]$ Y. Gong , B. Wang , A. Wang , {\it JCAP } {\bf 0701} 024 (2007);\\

$[14]$ M.R. Setare , {\it JCAP} {\bf 01} 023 (2007).\\

$[15]$ R. G. Cai and S. P. Kim, {\it JHEP} {\bf 02} 050 (2005).\\

$[16]$ M. Li , {\it Phys. Lett. B} {\bf 603} 01 (2004);\\

$[17]$ G. Izquierdo and D. Pavon, {\it Phys. Lett. B} {\bf 633} 420 (2006).\\

$[18]$ N. Mazumder and S. Chakraborty, {\it Accepted for
publication in Gen. Rel. Grav. in September} doi:10.1007/s10714-009-0881-z.\\\\

\end{document}